\documentclass[twocolumn,aps,prb,showpacs,preprintnumbers,amsmath,amssymb]{revtex4}
\usepackage{graphicx}
\usepackage{bm}
\usepackage{gensymb}
\usepackage{color}
\begin{document}
\title{ Chemical  control of polar behavior in bicomponent short-period superlattices}
\author{Hena Das$^1$, Nicola A. Spaldin$^2$, Umesh V. Waghmare$^3$ and T. Saha-Dasgupta$^1$}
\affiliation{$^1$ S.N. Bose National Centre for Basic Sciences,
Kolkata 700098, India}
\affiliation{$^2$ Materials Department, University of California, Santa Barbara, CA 93106-5050, USA}
 \affiliation{$^3$  Jawaharlal Nehru Centre for Advanced
Scientific Research,  Jakkur, Bangalore-560 064, India}
\date{\today}
\begin{abstract}
Using first-principles density functional calculations, we study 
the interplay of ferroelectricity and polar discontinuities in a range of 1-1 
oxide superlattices, built out of ferroelectric and paraelectric components. Studies have been carried
out for a varied choice of chemical composition of the components.
We find that, when polar interfaces are present, the polar discontinuities 
induce off-centric movements in the ferroelectric layers, even 
though the ferroelectric is only one unit cell thick. The 
distortions yield non-switchable polarizations, with magnitudes comparable to 
those of the corresponding bulk ferroelectrics. 
In contrast, in superlattices with no polar discontinuity at the interfaces, the off-centric movements in 
the ferroelectric layer are usually suppressed. The details of the behavior and
functional properties are, however, found to be sensitive to epitaxial strain, rotational instabilities 
and second-order Jahn-Teller activity, and 
are therefore strongly influenced by the chemical composition of the 
paraelectric layer. 

\end{abstract}
\maketitle

\marginparwidth 2.7in
\marginparsep 0.5in
\def\msm#1{\marginpar{\small MS: #1}}
\def\nsm#1{\marginpar{\small NS: #1}}
\def\tsd#1{\marginpar{\small TSD: #1}}
\def\scr{\scriptsize}
 \def\E{\mathcal{E}}

\section{Introduction}

Superlattices formed by layer-by-layer epitaxial growth of perovskite-based oxide materials are currently
a subject of intense research, because of their promising technological applications as well as 
fundamental scientific interest\cite{Ogale}. 
In ABO$_3$ perovskites, the A$^{+2}$B$^{+4}$O$_3$ (II-IV) 
structures consist of (100) layers of formally charge neutral AO and BO$_2$, while 
A$^{+3}$B$^{+3}$O$_3$ (III-III) 
or A$^{+1}$B$^{+5}$O$_3$ (I-V) structures have charged planes, composed of +1 AO and -1 BO$_2$ 
layers or -1 AO and +1 BO$_2$ layers, respectively. 
By stacking two perovskite layers from different charge families 
along the [001] direction, one obtains a {\it polar discontinuity} at the interface. 
Such polar discontinuities have been reported to lead to nontrivial local structural and electronic 
ground states\cite{LS1,mag,super,phase}, which are often not present in the parent bulk 
compounds \cite{Ogale}$^,$\cite{Altieri,Millis}. 
Investigating the properties of these ``exotic'' local phases has been an increasingly
active area of research in the past few years, particularly following a 2004 report\cite{LS1} 
of a conducting quasi-two dimensional electron gas (2DEG) at the interface 
between two wide-band insulators, LaAlO$_3$ (LAO) and SrTiO$_3$ (STO).
%
%
%
The measured mobility and carrier density of the LAO/STO interface are an
order of magnitude larger than those in analogous semiconductor-based systems\cite{semi}. 
Furthermore, magnetism, \cite{mag} superconductivity \cite{super} and a rich electronic 
phase diagram \cite{phase} have also been reported for this same system.
These fascinating and unexpected effects have generated strong excitement, and an intense
effort is currently devoted to better understanding the fundamental mechanisms of charge 
compensation at polar oxide-oxide interfaces. 
Parallel to this thrust, from the materials-design point of view, it is also important to 
identify new compounds, artificial superlattices or interfaces that might display similar (or 
possibly even more striking) behavior. 

A system can respond in several different ways to avoid a so-called 
{\it polar catastrophe}\cite{polar} -- a divergence in the potential 
caused by the polar discontinuity --
at such an interface between two charge-mismatched perovskites.
Compensation by free carriers was proposed in Ref.~\onlinecite{LS1}, and is
consistent with the observed conductivity at the interfaces. Other likely
possibilities are direct ionic charge compensation through mixed valency 
of the B cation, ion intermixing or oxygen vacancies at the interface\cite{LS2},
or polar distortions at the interface. These can be ``induced'' by the polar
discontinuity if both materials are paraelectric (PE) in their bulk phase
\cite{Pickett}, or ``natural'' if one or both components in the superlattice
is ferroelectric (FE)\cite{Vanderbilt}.
%
%
%
%

The mechanisms underlying induced and natural polar distortions can be readily
understood in terms of classical electrostatics
and the modern theory of polarization\cite{modern}, whenever the relevant
layers in the superlattice are at least three or four unit cells thick.
In particular, the charge mismatch can be interpreted as a polarization
discontinuity, which produces macroscopic electric fields in one or both 
components, because the normal component of the electric displacement field, 
$D=\E+4\pi P$, is preserved at a coherent insulating interface.
For smaller thicknesses, macroscopic concepts lose their meaning, as each
film is too thin to identify a well-defined local value of the electric field
or the polarization.
Thus, it remains an important question whether the above concepts 
are still valid when the layers in a superlattice are made thinner and 
thinner, down to the ultra-thin extreme limit of a single unit cell, or whether
new phenomena arise that radically alter the physics.
It is the main scope of this work to answer this question by studying 
some computer designed examples of 1-1 superlattices, constructed out of
ferroelectric and paraelectric layers, with different composition
and choices of polar/non-polar combinations.
We stress that this question has some points of contact with, but is 
largely unrelated to the better-known problem of the critical thickness 
for ferroelectricity\cite{oneunit} since the present problem deals with non-switchable
polar distortions driven by a polar discontinuity rather than switchable ferroelectric 
behavior.

In addition to issues related to the presence or absence of a polar discontinuity, 
other factors can influence or complicate the properties of
the 1-1 bicomponent superlattices. In particular, structural instabilities
such as the rotation and tilting of oxygen octahedra, the presence of misfit 
strain, and the presence of $d^{0}$ ions -- associated with a tendency to 
ferroelectricity -- in otherwise paraelectric layers, are likely to strongly 
affect the behavior.
It is therefore of interest to study how these additional factors may
alter the properties of the short-period superlattices compared to those
expected solely out of 
consideration of the presence or absence of a polar discontinuity.

%


To separate the role of electrostatics from the role of the specific choice 
of cations within a given charge family, we investigate
bicomponent superlattices with alternating III-III/II-IV and I-V/II-IV 
perovskite layers (with interfacial polar discontinuities), as well as others 
with alternating II-IV/II-IV and I-V/I-V perovskite layers (without polar 
discontinuities), both for a variety of chemical constituents. 
We find that our 1-1 structures behave in a way which is qualitatively analogous
to the longer-period superlattices previously investigated in the literature.
In particular, the systems containing polar discontinuities have strongly broken
inversion symmetry and large polarizations, but are not switchable. The magnitudes
of the polar distortions are as large as those in the corresponding bulk FE 
materials. The superlattices formed out of PE and FE layers with no potential 
discontinuity at the interface, on the other hand, usually have their off-centric 
displacements suppressed. We do, however, find examples of superlattices without polar 
mismatch that show marked ferroelectric instability, driven by either suppression 
of rotational instability or due to the presence of $d^0$ ions in the PE 
layer.

The remainder of the paper is organized as follows. Section II includes 
the methodology and the details of the computations carried out in this study. 
In Section III we present our results and discussion in several 
sub-sections dealing respectively with the bulk properties of the constituent materials, 
the calculated structural properties of the superlattices (with and without polar
discontinuities), and the calculated 
functional properties including static dielectric constants. Section IV provides the 
summary.

\section{Methodology and Computational Details}

Our calculations are performed using the plane wave based pseudopotential 
framework of density functional theory (DFT)\cite{11,12} as implemented in 
the Vienna ab initio Simulation Package (VASP)\cite{vasp1,vasp2}. 
We choose the local density approximation (LDA) rather than generalized 
gradient approximation (GGA) to describe the exchange-correlation functional 
due to the later's systematic overestimation of polar behavior\cite{13}. 
The optimized geometries are obtained by full relaxation of the atomic 
positions and out-of-plane lattice constants ($c$), with the in-plane 
lattice constants ($a$) fixed to the LDA bulk lattice constant of the 
ferroelectric component in the tetragonal phase. The positions 
of the ions are relaxed toward equilibrium until the 
Hellmann-Feynman forces become less than 0.001 eV/\AA. 
A 4$\times$4$\times$4 Monkhorst-Pack k-point mesh\cite{15} 
and 450 eV plane-wave cut-off give good convergence of the computed 
ground state properties.

The Born effective charges, phonon frequencies and dielectric tensors are obtained
from linear response calculations\cite{baroni}, using a variational formalism \cite{gonze} of
the density functional perturbation theory, as implemented in the ABINIT code. We 
calculate the electronic contribution to the polarization as a Berry phase using the
method developed by King-Smith and Vanderbilt \cite{kv}, and extract the ionic contribution
by summing the product of the position of each ion with its pseudo-charge.

The first part of our study concerns superlattices formed by alternating 
FE and PE single simple perovskite unit cells in the $z$ direction, and 
constraining them to be tetragonal in the $x-y$ plane. 
In a second stage, we consider structures formed by alternating
2$\times$2$\times$2 PE and FE cubic perovskite unit cells 
to allow rotation and tilting of the BO$_6$ octahedra. 

\begin{figure*}
\begin{center}
\rotatebox{-90}{\includegraphics[width=8cm,keepaspectratio]{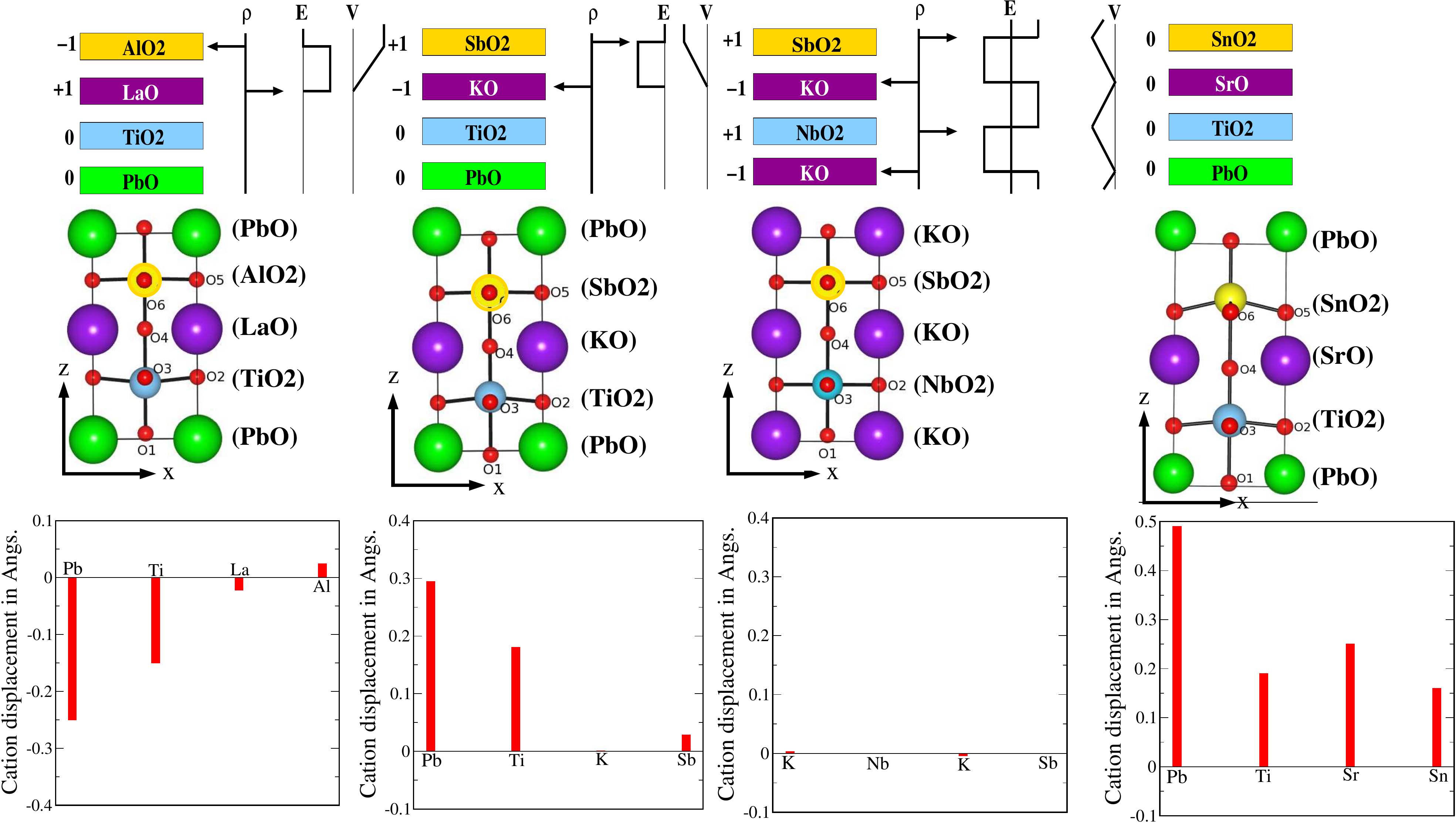}}
\caption{\label{fig-1}(Color online) Upper panel: Polar catastrophe for atomically abrupt (001) interfaces between LAO and PTO, KTO and PTO, KSO and KNO, SSO and PTO (left to right). The charges of each AO and BO$_2$ layer are taken from the formal charges of the constituent ions, after 
Ref.~\onlinecite{LS1}. 
Middle panel: Optimized structures for LAO/PTO, KTO/PTO, KSO/KNO and SSO/PTO, projected onto the
$xz$ plane. The large spheres indicate the cations at various layers, while the small spheres indicate oxygens. Inequivalent oxygens are indicated.
Lower panel: Cation displacements along the z axis, with respect to the centers of their 
oxygen cages.} 
\end{center}
\end{figure*} 

\section{Results and Discussion}

\begin{table} 
\caption{ \label{systems} Representative bicomponent superlattices composed of 
paraelectric (PE) and ferroelectric (FE) components.}
\begin{tabular}{|c|c|c|c|c|c|c|}
\hline
 & PE & FE & Polar Interface? & Mismatch (\%)\\
\hline
 LAO/PTO & LaAlO$_3$ & PbTiO$_3$ & YES & 0.99 \\
\hline
 KSO/PTO & KSbO$_3$  & PbTiO$_3$ & YES & 1.94 \\
\hline
 KSO/KNO & KSbO$_3$  & KNbO$_3$  & NO & 0.10 \\
\hline
 SSO/PTO & SrSnO$_3$ & PbTiO$_3$ & NO & 4.40 \\ \hline
\end{tabular}
\end{table}

We consider the four representative systems listed in Table~\ref{systems}. All
combinations contain one material which is ferroelectric in the bulk, and one
which is paraelectric. 
Two of the systems (LAO/PTO and KSO/PTO) have polar discontinuities at the interface
whereas the other two (KSO/KNO and SSO/PTO) do not. 
We deliberately choose paraelectric components without $d^0$ cations, because $d^0$
cations have a strong tendency to off-center through the so-called second-order
Jahn-Teller effect. This often results in ferroelectricity, and in materials with
paraelectric ground states it can cause incipient or quantum ferroelectricity\cite{qfe},
which we choose to avoid here.
We begin by briefly summarizing
the relevant bulk properties of the various constituents. 

\subsubsection{Bulk properties}

Both  PTO and KNO are well-known ferroelectric materials with the cubic perovskite structure at high 
temperature. PTO undergoes a phase transition to a tetragonal phase below 766 K, with the polarization 
along the [001] direction. In the tetragonal phase the experimental in-plane ($a$) and out-of plane 
($c$) lattice parameters are 3.90 \AA\ and 4.14 \AA\ respectively\cite{PTO} and the ferroelectric
polarization is 59 $\mu$C/cm$^{-2}$; our calculated corresponding LDA values 
are 3.85 \AA\ and 4.05  \AA, and 80.45 $\mu$C/cm$^{-2}$ respectively. 
Bulk KNO is a rhombohedral ferroelectric with a polarization of 42  $\mu$C/cm$^{-2}$ along [111] 
below its Curie temperature\cite{kno-expt} of 230 K. Earlier theoretical work reported stabilization of the
tetragonal phase with compressive strain \cite{KNO}; for tetragonal KNO we obtain in- and out-of-plane 
lattice constants of 3.92 \AA\ and 4.00 \AA\ and a polarization of 26 $\mu$C/cm$^2$ along [001]. 

LAO is a wide band gap insulator with strong alternating rotations of the oxygen octahedra around the [111]
direction leading to rhombohedral symmetry \cite{LAO}. Recent density functional calculations 
showed that epitaxial strain causes a polar instability which competes with these non-polar 
oxygen rotations\cite{Hatt-Spaldin:2009}. The lattice parameter for the cubic structure
without oxygen rotations is 3.81 \AA \cite{LAO1}. KSO is not known experimentally. Computationally
we obtain a cubic LDA optimized lattice constant of 3.92 \AA. 
At room temperature SSO crystallizes in the centrosymmetric orthorhombic space group $Pbnm$\cite{SSO}, 
with lattice parameters $a= 5.71$ \AA, $b= 5.71$ \AA\ and $c= 8.06$ \AA. The structure is characterized by 
a classic GdFeO$_3$ tilt pattern of the SnO$_6$ octahedra. A structural transition from the ground-state
orthorhombic structure to a high temperature tetragonal $I4/mcm$ phase has been found\cite{sso2} 
driven by an order-disorder octahedral tilting transition.

\subsubsection{Structural properties}

We begin by optimizing the geometries of our four representative superlattices. We 
construct the superlattices by alternating FE and PE layers along the [001] direction.
We impose an in-plane periodicity of one simple cubic perovskite unit cell: This choice
imposes an overall tetragonal symmetry and prohibits rotations and tiltings of the
oxygen octahedra. We investigate the effects of relaxing this constraint later. 
By fixing the in-plane lattice constant to the LDA value for the FE layer we induce
a small tensile strain in the PE layer for LAO on PTO, small compressive strains 
for KSO on PTO or KNO, and a large compressive strain for SSO on PTO 
(values in Table~\ref{systems}).
We then fully relax the out-of-plane lattice parameter and the ionic positions within 
the tetragonal symmetry. All superlattices are found to be insulating within the LDA with band 
gaps of $\sim$2 eV, allowing us to directly calculate their ferroelectric polarization
using the Berry phase formalism \cite{kv}.

Both superlattices containing polar discontinuities
are strongly polar, with polarizations of -50.87 $\mu$C/cm$^{2}$ for LAO/PTO  
and +57.74 $\mu$C/cm$^{2}$ for KSO/PTO; the opposite signs in the two cases
reflect the opposite orientation of the polar discontinuity (III-III/II-IV
in the first case and I-V/II-IV in the second). 
Neither case is ferroelectric, because the polarization orientations are fixed 
by the polar discontinuity and the polarizations are not switchable \cite{Vanderbilt}.
The calculated structures are shown in Fig.1.
In LAO/PTO, the Pb and Ti ions displace from the centers of their coordination 
polyhedra by $\approx$0.25 \AA\ and 0.15 \AA\ respectively along the
positive crystallographic {\it z} axis, that is towards the PbO/AlO$_2$ interface.
In KSO/PTO, Pb and Ti off-center by 0.29 \AA\  and 0.18 \AA\ respectively
along the negative {\it z} axis, away from the PbO/SbO$_2$ interface. 
The magnitudes of these displacements are similar to those in bulk PTO, which are 0.41 \AA\ and 0.26 \AA\  for Pb and Ti respectively.
Note that the off-centric movements in the LaO/KO and AlO$_2$/SbO$_2$
layers are tiny, with the movements of K and Sb being in the same direction 
as Pb and Ti, and La and Al moving in the opposite direction to each other. 
We, however, do not attach much significance to these small displacements. 

The situation for the superlattices without polar discontinuities is  
more complicated. For KSO/KNO we indeed obtain a negligible polarization of 
0.02 $\mu$C/cm$^{2}$, resulting from the tiny displacements of K and Nb from the centers of their 
coordination polyhedra\cite{footnote}. Note that the KSO/KNO superlattice in its paraelectric phase 
has an inversion center. However we do not impose the inversion symmetry during the relaxation procedure, 
since our aim is to find out the existence or non-existence of the polar behavior. The tiny polarization
obtained for relaxed geometry of KSO/KNO is within the accuracy limit of our calculations. This leads us to conclude that 
our obtained geometry is in fact centrosymmetric, as has been checked in terms of the energy difference
between the fully centrosymmetric case and the relaxed structure, which turned out to be less than
0.01 meV.
%
For SSO/PTO, however, we obtain a rather large polarization 
of $\sim$ -66 $\mu$C/cm$^2$, with a cooperative displacement of all cations
(Sr, Sn, Pb, Ti). Interestingly, the polarization is found to be 
switchable: In Fig. 2 we show the calculated total energy, $E$ as a function of 
polarization $P$ obtained by calculating polarizations and total energies of 
geometries with fixed values of off-centric movements. The fixed values of off-centric movements
are obtained by interpolating between off-centric movements of ions corresponding to two minima 
of the $E$ versus $P$ curve.
The $E(P)$ curve shows the characteristic double well structure typical of 
switchable ferroelectrics, except that in this case it is strongly asymmetric as a
result of the additional inversion-center lifting that is built in to the structure 
through the layering. 
The switchable behavior indicates that the polarization arises from an origin 
other than the polar discontinuity alone, as the polar discontinuity can only 
drive the polarization in one direction. 
Note that it is also likely distinct from the behavior in
BaTiO$_3$/SrTiO$_3$ superlattices\cite{bto-sto} where the PE component (STO)
is a quantum ferroelectric.  In the next section we explore the origin of this 
behavior. 

\begin{figure}
\begin{center}
\includegraphics[width=6cm,keepaspectratio,angle=-90]{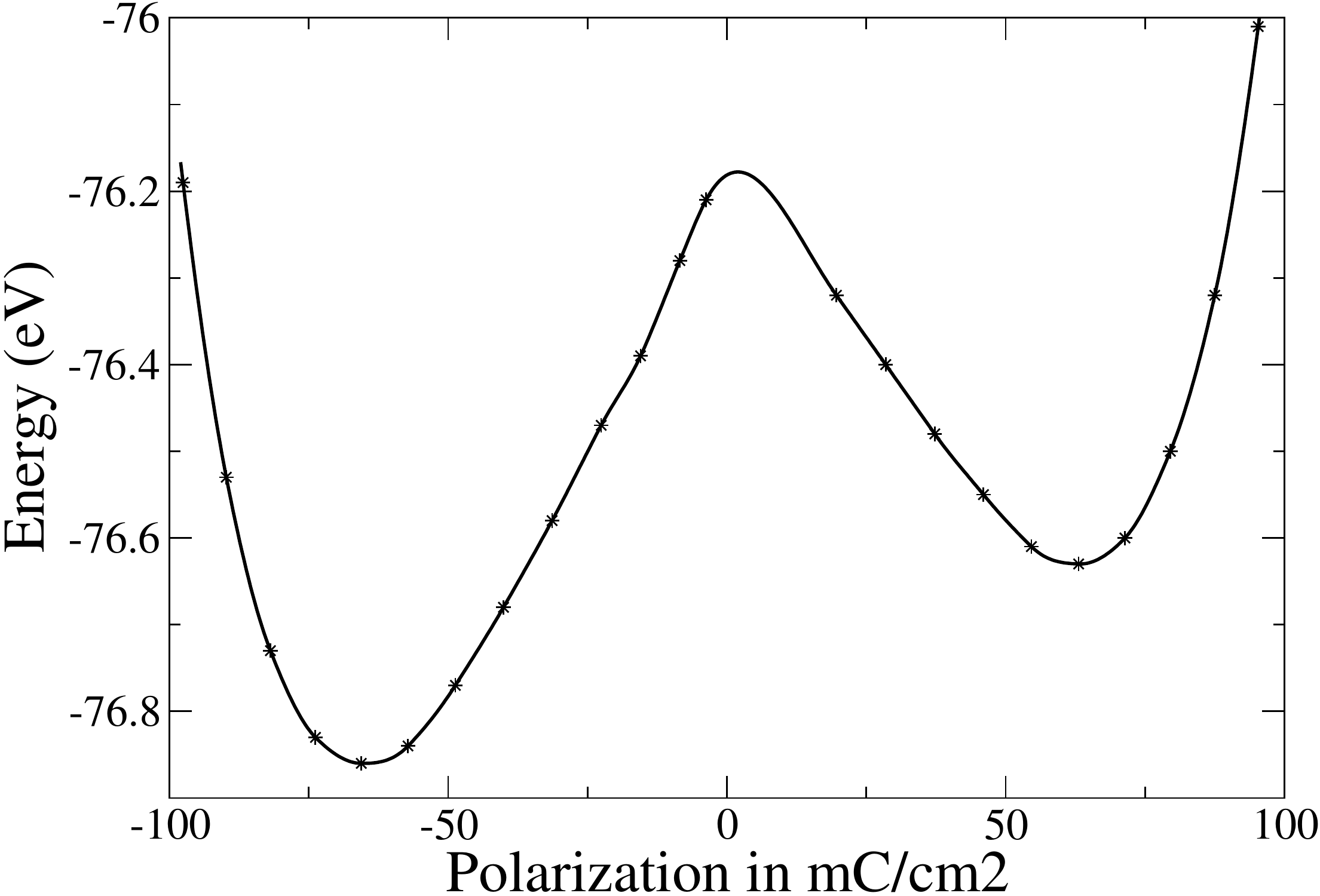}
\caption{\label{fig-2}Potential energy curve of the total polarization for SSO/PTO superstructure for 1$\times$1$\times$1/1$\times$1$\times$1 geometry (see 
text for details).}
\end{center}
\end{figure}

\subsubsection{Origin of polarization at non-polar interfaces}

Our first clue as to the origin of the unexpected polarization in 
the SSO/PTO superlattice lies in the large (4.4\%) mismatch in lattice 
constants between the two constituents in this system. 
To test the role of
the misfit strain in driving polarization, we performed an analogous set 
of calculations for SrSnO$_3$/BaTiO$_3$ (SSO/BTO), in which the misfit 
strain (estimated using the
experimental bulk lattice constants \cite{23,17}) is only 0.78 \%. Indeed, we find
that the calculated polarization for SSO/BTO is considerably reduced 
(to -7.76 $\mu$C/cm$^2$) compared with SSO/PTO, 
and the corresponding off-center displacements are 0.02 \AA, 0.06 \AA, 0.06 \AA\ and 
0.02 \AA\ for Sr, Sn, 
Ba and Ti respectively. We have further carried out calculations for bulk SrSnO$_{3}$ 
in tetragonal geometry, disabling the rotation and tilt of the SnO$_6$ octahedra, under
4.4 \% and 0.78 \% compressive strains (strains identical to SSO/PTO and SSO/BTO
systems respectively). These calculations resulted in large off-centric displacements
of Sr$^{2+}$ and Sn$^{4+}$ ions for 4.4 \% compression strain and negligible displacements
for 0.78 \% strain value, illustrating once more the essential role of the strain in determining 
the polarization. 
%
%

While it is an already established result \cite{strain} that strain plays a crucial 
role in determining polarization and off-centering in superlattices, there has also
been recent discussion of competition between tilts and rotations of the oxygen
octahedra and ferroelectric displacements. 
For example it has been shown computationally\cite{Hatt-Spaldin:2009} that LaAlO$_3$
shows ferroelectric behavior under epitaxial strain when the octahedral rotations 
are suppressed, while ferroelectricity is suppressed when those
rotations are allowed to relax \cite{Hatt-Spaldin:2009}.
Similar physics led to predictions of ferroelectricity in A-site alloy perovskites 
\cite{singh08}, in which the polar behavior
is obtained by destabilizing the cooperative rotational instability through A-site disorder.
And in Ref.~\onlinecite{camno}, polar behavior
was engineered theoretically in CaMnO$_3$ by suppressing the orthorhombic tilting instability
through strain and chemical substitution. The possibility of competing rotational
instability governed by zone-boundary phonons with ferroelectric modes should
therefore be carefully analyzed before drawing a definitive conclusion on a specific
system. 
To check for this possibility in our superlattices, we 
increased the size of our supercell to 2 $\times$ 2 $\times$ 2 simple
perovskite unit cubes in order to allow for oxygen rotations and tiltings
and re-compute the lowest energy structures.
Indeed
we find that, while for LAO/PTO, KSO/PTO and KSO/KNO our conclusions remain
unchanged, for SSO/PTO the system adopts a centrosymmetric structure with 
negligible off-centering when rotational distortions are allowed. Bulk SrSnO$_3$ calculations
allowing for the octahedral rotation and tilt, as expected, also yielded a centrosymmetric
structure. We find applied compressive strain increases the rotation and decreases the tilt
in agreement with the findings in Ref~\onlinecite{Hatt-Spaldin:2009}.
This reinforces the earlier suggestions in the literature, and implies that 
if octahedral rotation at the SSO layer can be disabled, which might be possible
in ultra-thin films using strain or doping, then SSO/PTO superlattices could 
have switchable polar properties.

\subsubsection{Ferroelectric tendency in the paraelectric layer}

In all of the model superlattices studied above, the paraelectric material was
chosen to be one that shows no tendency to ferroelectric off-centering in its
bulk phase. In this section we study the effect of substituting it with an
``incipient ferroelectric'' or ``quantum ferroelectric'', i.e. a material 
that is still PE in the bulk, but very close to a ferroelectric phase transition.

Specifically we compare our results reported above for KSO/KNO with analogous
calculations for KTaO$_3$/KNbO$_3$ 1-1 superlattices. We choose KTaO$_3$ to
compare with KSbO$_3$ because the former has a so-called Second Order Jahn-Teller
(SOJT) active\cite{SOJT} ion with a $d^0$ electron configuration which is associated
with a tendency to off-center. Indeed bulk KTO is a quantum ferroelectric with
ferroelectricity suppressed by quantum fluctuations\cite{qfe}. 
Ultrashort-period (KTO)$_m$/KNO superlattices have been previously 
investigated\cite{KTO} in terms of ab-initio studies.
As in Ref.~\onlinecite{KTO} for $m$ = 1, we find that KTO/KNO is strongly ferroelectric, with 
a switchable polarization of 16.46 $\mu$C/cm$^2$ and atomic 
displacement at the Ta site of 0.14 \AA.

The difference in physical behavior between KTO/KNO and KSO/KNO must arise from  
the chemical difference between KTO and KSO. To further understand this, we next
compute the Born effective charges, 
Z$^*$ for KTO and KSO, using the {\it c/a} ratios that we obtained 
for the superlattice geometries. The Ta Z$^*$ in KTO is 
highly anomalous (8.78 instead of the nominal valence of 5), consistent
with the proximity of the $d^0$ KTO system to ferroelectricity,  
while the Sb Z$^*$ in KSO is 5.4, close to the nominal value of 5. 
The strong ferroelectric instability in KNO is therefore able to induce
a ferroelectric instability in the proximally ferroelectric KTO layer;
KSO is less polarizable and therefore resists polar distortion.

We also compute the properties of the KTaO$_3$/PbTiO$_3$ superlattice in order
to study the behavior of a ferroelectrically active paraelectric in 
a superlattice with a polar discontinuity. Here KTO is under 3.51 \% compressive 
strain. We find a non-switchable polarization of 69.72 $\mu$C/cm$^2$ and displacements of 0.39 \AA, 
0.26 \AA, 0.06 \AA\ and 0.10 \AA\ for Pb, Ti, K and Ta respectively; all of these
quantities are larger than the corresponding values for LAO/PTO and KSO/PTO, 
expected because in this case SOJT activity is additive with
polar discontinuity behavior.

\subsubsection{Static Dielectric constant}
 
Finally, for completeness, we compute the dielectric properties of the polar PE-FE
superlattices driven by polarization discontinuity, specifically LAO/PTO and KSO/PTO.
Our motivation lies in the fact that the dielectric constants typically hold signatures 
of polar instability of soft modes\cite{UVW} and are also of technological importance.

The $\alpha\beta$ component of the static dielectric constant is given by,
\begin{equation}
\epsilon_{\alpha\beta} = (\epsilon^{\varpropto})_{\alpha\beta} +\dfrac{4\pi}{\Omega_0} \sum_{\omega^2_{\mu}\neq0}\dfrac{\bar{Z}^*_{\mu\alpha}\bar{Z}^*_{\mu\beta}}{\omega^2_{\mu}}
\end{equation}
where $\epsilon^{\varpropto}$ is the electronic contribution to the dielectric tensor and the 
second term is the sum of the contributions from each polar phonon. $\mu$ labels the phonon modes with frequency 
$\omega_{\mu}$, $\bar{Z}^{*}_{\mu\alpha}$ is the mode effective charge corresponding to mode $\mu$ in the Cartesian direction $\alpha$ and 
$\Omega_{0}$ is the volume of the primitive unit cell.
The mode effective charge in the $\alpha$ direction for a given mode $\mu$ is related to the eigendisplacement $U_{\mu}(\kappa\beta)$
involving ion $\kappa$ and the Born effective charge tensor, Z$^{*}$ by,
\begin{equation}
\bar{Z}^*_{\mu ,\alpha}=\dfrac{\sum_{\kappa\beta}Z^*_{\kappa ,\alpha\beta}U_{\mu}(\kappa\beta)}{\left[\sum_{\kappa\beta}U^*_{\mu}(\kappa\beta)  U_{\mu} (\kappa\beta)\right]^{1/2}}  .
\end{equation}

{\it Electronic dielectric constant:} The results for the electronic dielectric tensor are shown in Table-II. The electronic contributions are of almost the same magnitude for LAO/PTO and KSO/PTO superlattices as expected from the similarity of their LDA band gaps.

\begin{table}
\caption{Electronic and lattice contribution of the static dielectric constant for LAO/PTO and KSO/PTO superlattices.}
\begin{tabular}{|c|cc|cc|}
\hline
&\multicolumn{2}{|c|}{Electronic contribution}&\multicolumn{2}{|c|}{Lattice contribution}\\  
&LAO/PTO&KSO/PTO&LAO/PTO&KSO/PTO\\
\hline
$\epsilon_{xx}$&6.39&5.93&137.19&90.12\\
$\epsilon_{yy}$&6.39&5.93&137.19&90.12\\
$\epsilon_{zz}$&6.14&5.34&43.09&24.99\\
\hline

\end{tabular}
\end{table}

{\it Lattice dielectric constant:} Our calculated lattice contributions to the dielectric tensor are 
listed in Table-II. We find that the lattice contribution to the static dielectric constant 
is quite different between LAO/PTO and KSO/PTO superlattices. To understand the origin of the difference, 
we next carry out detailed analysis of the phonon frequencies and mode effective charges for the phonon modes 
which contribute to $\epsilon_{zz}$; that is the relevant quantity for superlattices grown along the 
[001] direction with significant off-centric movements along the z-axis. A large lattice dielectric response can be the result of
the presence of one or more very low frequency polar phonons and/or anomalously large mode effective charges.
Our analysis, as listed in Table-III, shows that for the LAO/PTO superlattice, the lattice dielectric constant originates essentially from the dominant contribution 
of the lowest mode at 115 cm$^{-1}$; this in turn arises from the combined effect of large mode 
effective charge and low frequency of the phonon mode. The contributions of different ions to the 
mode effective charge are found to be of comparable magnitudes.

In the case of the KSO/PTO superlattice, on the other hand, the dominant contributions come from 
more than one phonon mode. The major contributions arise from the phonon mode at 
209.94 cm$^{-1}$, which has a giant mode effective charge, and that at 125.83 cm$^{-1}$, which has low frequency. 
The giant mode effective charge for the phonon mode at 209.94 cm$^{-1}$ 
comes primarily from the Ti$^{+4}$ and Sb$^{+5}$ ions and the planar oxygens situated at the SbO$_2$ layer.
Thus, the temperature dependence of dielectric response of
 KSO/PTO superlattice is expected to be weaker than that of
the LAO/PTO superlattice.

\begin{table}
\caption{Phonon frequencies in cm$^{-1}$, corresponding mode effective charges in unit of $\vert e \vert$ and the contribution to 
static dielectric constant.}
\begin{tabular}{|ccc|ccc|}
\hline
\multicolumn{3}{|c|}{LAO/PTO}&\multicolumn{3}{|c|}{KSO/PTO}\\  
\hline
$\omega$&$\bar{Z}^{*}_{\mu,zz}$&$\epsilon^{zz}_{\mu}$&$\omega$&$\bar{Z}^{*}_{\mu,zz}$&$\epsilon^{zz}_{\mu}$\\
\hline
115.19&0.82&21.95&125.83&0.57&8.32\\
181.17&0.70&6.50&194.37&0.60&4.03\\
302.57&1.24&7.33&209.94&1.06&10.06\\
363.04&0.74&1.83&284.69&0.36&0.68\\
564.96&0.60&0.49&458.66&0.43&0.37\\
681.35&1.01&0.95&639.09&1.08&1.20\\
757.08&0.27&0.05&801.07&0.84&0.46\\
\hline
\end{tabular}
\end{table}

\section{Summary}
We have carried out a detailed computational study of superlattices formed out of alternating
layers of one unit cell thick FE and PE components, considering several different ABO$_{3}$ 
perovskites. We find that the tendency of the superlattices to exhibit polar properties depends 
strongly on  the chemistry of the PE and FE components. In addition to interfacial polar 
discontinuities at the interfaces between the PE and FE layers, the epitaxial strain, rotational 
instability of the PE layer and the presence of $d^0$ second order Jahn-Teller active ions in 
the PE layer contribute to determining the polar behavior. While the polarization arising from a 
polar discontinuity is generally not switchable, other routes to polar behavior discussed here, 
do allow polarization switching.
The strongest polar response was found for superlattices 
with polar discontinuities and formed out of FE and $d^0$ PE layers. 

\section{Acknowledgment}
The work was supported by National Science Foundation under Award Nos. DMR-0940420 (NAS) and
DMR-0843934 (the International Center for Materials Research). TSD acknowledges the support of Department of 
Science and Technology through Advanced Materials Research Unit and Swarnajayanti Research grant. UVW acknowledges 
partial support for this work from IBM faculty award. The authors gratefully acknowledge the contribution of M. Stengel 
in terms of the initiation of this work and for several useful discussions.


\begin{thebibliography}{99}
\bibitem{Ogale} J. Mannhart, in Thin Films and Heterostructures for Oxide Electronics, S. Ogale, Ed. (Springer, New York, 2005), pp. 251-278.
\bibitem{LS1} A. Ohtomo and H. Y. Hwang, Nature {\bf 427}, 423 (2004).
\bibitem{mag} A. Brinkman, M. Huijben, M. van Zalk, J. Huijben, U. Zeitler, J. C. Maan, W. G. van der Wiel, G. Rijnders, D. H. A. Blank, H. Hilgenkamp, Nature Mater. {\bf 6}, 493 (2007).
\bibitem{super}N. Reyren, S. Thiel, A. D. Caviglia, L. Fitting Kourkoutis, G. Hammerl, C. Richter, C. W. Schneider, T. Kopp, A.-S. RÃ¼etschi, D. Jaccard, M. Gabay, D. A. Muller, J.-M. Triscone, and J. Mannhart, Science {\bf 317}, 1196 (2007).
\bibitem{phase}A. D. Caviglia, S. Gariglio, N. Reyren, D. Jaccard, T. Schneider, M. Gabay, S. Thiel, G. Hammerl, J. Mannhart, J.-M. Triscone, Nature {\bf 456}, 624 (2008).
\bibitem{Altieri}S. Altieri, L. H. Tjeng, G. A. Sawatzky, Thin Solid Films {\bf 400}, 9 (2001).
\bibitem{Millis} S. Okamoto and A. J. Millis, Nature {\bf 428}, 630 (2004).
\bibitem{semi} G. A. Baraff, J. A. Appelbaum and D. R. Hamann, Phys. Rev. Lett. {\bf 38} 237 (1977); W. A. Harrison, E. A. Kraut, J. R. Waldrop and R. W. Grant, Phys. Rev. B {\bf 18} 4402 (1978).
\bibitem{polar} C. Noguera, J. Phys.: Condens. Mater. {\bf 12} R367 (2000).
\bibitem{LS2} N. Nakagawa, H. Y. Hwang and D. A. Muller, Nat. Mater. {\bf 5}, 204 (2006).
\bibitem{Pickett} R. Pentcheva and W. Pickett, Phys. Rev. Lett. {\bf 102}, 107602 (2009).
\bibitem{Vanderbilt} E. Murray and D. Vanderbilt, Phys. Rev. B {\bf 79}, 100102 (2009).
\bibitem{modern} R. Resta and D. Vanderbilt, "Theory of Polarization: A Modern Approach," in {\it Physics of Ferroelectrics: a Modern Perspective}, ed. by K.M. Rabe, C.H. Ahn, and J.-M. Triscone (Springer-Verlag, 2007, Berlin), pp. 31-68. 
\bibitem{oneunit} N. A. Spaldin, Science {\bf 304} 1606 (2004); D. D. Fong {\it et.al.} Science {\bf 304} 1650 (2004).
\bibitem{qfe} U. T. H\"{o}chli and L. A. Boatner, Phys. Rev. B {\bf 20}, 266 (1979).  
\bibitem{11} P. Hohenberg and W. Kohn, Phys. Rev. {\bf 136}, B864 (1964).
\bibitem{12} W. Kohn and L. J. Sham, Phys. Rev. {\bf 140}, 1133A (1965).
\bibitem{vasp1} G. Kresse and J. Hafner, Phys. Rev. B {\bf 47}, R558 (1993).
\bibitem{vasp2} G. Kresse and J. Furthmuller, Phys. Rev. B {\bf 54}, 11169 (1996).
\bibitem{13} Y. Umeno, B. Meyer, C. ElsÃ¤sser, and P. Gumbsch, Phys. Rev. B {\bf 74}, 060101(R) (2006).
\bibitem{15} H.J.Monkhorst and J. D. Pack, Phys. Rev. B {\bf 13}, 5188 (1976).
\bibitem{baroni} S. Baroni, P. Giannozzi and A. Testa, Phys. Rev. Lett {\bf 58} 1861 (1987).
\bibitem{gonze} X. Gonze, D. C. Allen and M. P. Teter, Phys. Rev. Lett {\bf 68} 3603 (1992);
 X. Gonze, Phys. Rev. B {\bf 55} 10337 (1997); X. Gonze and Ch. Lee, Phys. Rev. B {\bf 55} 10355 (1997).
\bibitem{kv} R.D.King-Smith and D. Vanderbilt, Phys. Rev. B {\bf 47}, 1651 (1993);  
D. Vanderbilt and R. D. King-Smith, Phys. Rev. B {\bf 48}, 4442 (1994).
\bibitem{PTO} G. Shirane, R. Pepinsky and B. C. Frazer, Acta. Crys. {\bf 9}, 131-140 (1956); A. Sani, M. Hanfland and D. Levy, J. Solid State Chem. {\bf 167}, 446â452 (2002).
\bibitem{kno-expt} A. W. Hewat, J. Phys. C {\bf 6} 1973 (1973).
\bibitem{KNO} O. Dieguez, K. M. Rabe and D. Vanderbilt, Phys. Rev. B {\bf 72}, 144101 (2005).
\bibitem{LAO} C. J. Howard, B. J. Kennedy, B. C. Chakoumakos, J. Phys.: Condensed Matter {\bf 12}, 349 (2000).
\bibitem{Hatt-Spaldin:2009} A. J. Hatt and N. A. Spaldin, arXiv:0808.3792v1 (2008)
\bibitem{LAO1} C. Howard, J. Phys.: Condenced Matter {\bf 12}, 349-365 (2000).
\bibitem{SSO} A. Vegas, M. Vallet-RegÄ±, J.M. Gonzales-Calbet, M.A. Alario-Franco, Acta Crystallogr. B {\bf 42} (1986), 167â172.
\bibitem{sso2} E. H. Mountstevens and S. A. T. Redfern, Phys. Rev. B {\bf 71}, 220102(R) (2005).
\bibitem{footnote} Polarization P, is a multi-valued function, it is only defined modulo a ``quantum of polarization'' $\Delta P$ = $e/S$, where $S$ is the cell surface area. From the raw polarization, as obtained from the Berry phase calculations, suitable number of
polarization quanta were, therefore, subtracted to arrive to the effective polarization values, defined relative to a centrosymmetric
reference. In case of systems like KSO/KNO consisting of charged AO and BO$_2$ layers, an extra half quantum was subtracted, as
explained in Ref. M. Stengel and D. Vanderbilt, Phys. Rev. B {\bf 80}, 241103 (R) (2009). 
\bibitem{bto-sto} J. B. Neaton and K. M. Rabe, Appl. Phys. Lett {\bf 82}, 1586 (2003).
\bibitem{23} H. D. Megaw, Proc. Phys. Soc., London {\bf 58}, 133-152 (1946).
\bibitem{17} H. T. Evans jr. , Acta. Crystallogr. {\bf 14}, 1019-1026 (1961).
\bibitem{strain} M. P. Warusawithana, C. Cen, C. R. Sleasman, J. C. Woicik, Y. Li, L. F. Kourkoutis, J. A. Klug, H. Li, P. Ryan, L. Wang, M. Bedzyk, D. A. Muller, L. Chen, J. Levy, and D. G. Schlom, Science {\bf 324}, 367 (2009); C. Fennie and K. M. Rabe, Phys. Rev. Lett {\bf 97}, 267602 (2006); E. Bousquet, N. A. Spaldin and P. Ghosez, Phys. Rev. Lett {\bf 104}, 037601 (2010).
\bibitem{singh08} D. J. Singh and C. H. Park, Phys. Rev. Lett. {\bf 100}, 087601 (2008).
\bibitem{camno} S. Bhattacharjee, E. Bousquet and P. Ghosez, Phys. Rev. Lett. {\bf 102}, 117602 (2009).
\bibitem{SOJT}I. Bersuker, Chem. Rev. (Washington, D. C.) {\bf 101}, 1067 (2001).
\bibitem{KTO} S. Hao, G. Zhou, X. Wang, J. Wu, W. Duan and B. Gu, App. Phys. Lett {\bf 86} 232903 (2005).
\bibitem{UVW} U. V. Waghmare and K. M. Rabe, 
{\it Materials Fundamentals of Gate Dielectrics}, pg 215-247, ed. A. A. Demkov and
A. Navrotsky, Springer (2005).



\end{thebibliography}
\end{document}